\documentclass[aps,prl,showpacs,twocolumn,superscriptaddress]{revtex4-1}
\usepackage[latin9]{inputenc}
\usepackage{textcomp}
\usepackage{amsmath}
\usepackage{amssymb}
\usepackage{epsfig}
\usepackage{epstopdf}
\usepackage{bm}
\usepackage{graphicx,epsfig}
\usepackage{mathrsfs}
\usepackage{dcolumn}
\usepackage{color}
\usepackage{natbib}
\usepackage{CJK}
\usepackage{miniplot}
\usepackage{subfigure}
\usepackage{graphicx}

\makeatletter

\newcommand*\LyXThinSpace{\,\hspace{0pt}}

\makeatletter

\newcommand{\Rmnum}[1]{\expandafter\@slowromancap\romannumeral #1@}
\makeatother

\usepackage{graphicx}
\usepackage{amsfonts}

\makeatother

\begin{document}

\title{Electromagnetically Induced Transparency with
Superradiant and Subradiant states}

\author{Wei Feng}
\affiliation{Beijing Computational Science Research Center, Beijing 100193,
China}
\affiliation{Texas A$\&$M University, College Station, TX 77843, USA}
\author{Da-Wei Wang }
\email{whatarewe@tamu.edu}
\affiliation{Texas A$\&$M University, College Station, TX 77843, USA}
\author{Han Cai}
\affiliation{Texas A$\&$M University, College Station, TX 77843, USA}
\author{Shi-Yao Zhu}
\affiliation{Beijing Computational Science Research Center, Beijing 100193,
China}
\affiliation{Department of Physics, Zhejiang University, Hangzhou 310027, China}
\author{Marlan O. Scully}
\affiliation{Texas A$\&$M University, College Station, TX 77843, USA}
\affiliation{Baylor University, Waco, Texas 76706, USA}
\affiliation{Xi'an Jiaotong University, Xi'an, Shaanxi 710048, China}

\date{\today }
\begin{abstract}
We construct the electromagnetically induced transparency (EIT) by dynamically coupling a superradiant state with a subradiant state. The superradiant
and subradiant states with enhanced and inhibited decay rates act as the excited and metastable states in EIT, respectively. Their energy difference
determined by the distance between the atoms can be measured by the EIT spectra, which renders this method useful in subwavelength metrology.
The scheme can also be applied to many atoms in nuclear quantum optics, where the transparency point due to counter-rotating wave terms can be observed.
\end{abstract}


\maketitle
\emph{Introduction}.--Electromagnetically induced transparency
(EIT) \cite{Harris1991,FleischhauerRMP} is a quantum optical mechanism that is
responsible for important
phenomena such as slow light \cite{Hau1999,Kocharovskaya2001,Kash1999},
quantum memory \cite{Lukin2000PRL,Lukin2002PRA,Lukin} and enhanced
nonlinearity \cite{Harris1990,Jain1996}. A probe field that resonantly
couples the transition from the ground state $|g\rangle$ to an excited
state $|e\rangle$ of an atom, experiences a transparency point at
the original Lorentzian absorption peak, if the excited state is coherently
and resonantly coupled to a metastable state $|m\rangle$. EIT involves
at least three levels and naturally three-level atoms
are used in most cases. However, proper three-level structures are not available
in some optical systems, such as in atomic nuclei
\cite{Rohlsberger2010,Anisimov2007,Tittonen1992}
and biological fluorescent molecules \cite{Bates2005}, in which
EIT can have important applications once realized. Interestingly, it has been shown that
even with only two-level systems, EIT-like spectra can
be achieved by locally addressing the atomic ensembles
\cite{Rohlsberger2012, Xu2013, Makarov2015}.
However, strict EIT scheme with a dynamic coupling field is still absent in two-level optical systems.

Superradiance and subradiance are the enhanced and inhibited
collective radiation of many atoms
\cite{Dicke1954,Lehmberg1970,Agarwal1974}, associated with the collective Lamb shifts \cite{Scully2009PRL,Scully 07
LaserPhysics,Dawei2010PRA}. The superradiance and subradiance of two interacting atoms has attracted much interest both theoretically \cite{Petrosyan2002, Muthukrishnan2004} and experimentally \cite{Grangier1985, DeVoe1996, Hettich2002, Gaetan2009, McGuyer2015}.
In this Letter, we use superradiance and subradiance to construct
EIT and investigate the new feature in the EIT absorption spectrum involving with the cooperative effect and the counter-rotating wave terms. For only two
atoms, the symmetric (superradiant) state has much larger
decay rate than the anti-symmetric (subradiant) state when the distance between the two atoms
is much smaller than the transition wavelength. These two states serve as the excited and the metastable states and their splitting, depending on the distance between the atoms, can be measured by the EIT spectra.
In addition, the counter-rotating wave terms in the effective coupling field between the superradiant and subradiant states bring an additional transparency point, which is usually not achievable in traditional EIT systems with three-level atoms.

\begin{figure}
\begin{centering}
\includegraphics[width=9cm]{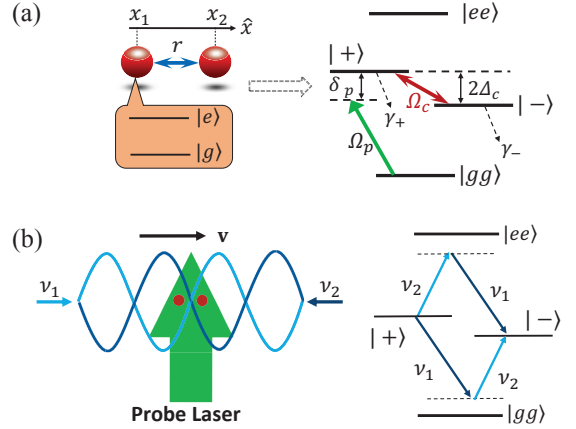}
\par\end{centering}

\caption{(Color online) (a) Two two-level atoms form an EIT system with the
symmetric (superradiant) state being the excited state and the anti-symmetric
(subradiant) state being the metastable state. (b) The symmetric and anti-symmetric
states are resonantly coupled by the Raman transitions of two counter-propagating plane waves. We can also understand this coupling as induced by the time-dependent difference between the dynamic Stark shifts of the two atoms induced by a moving
standing wave with velocity $\mathbf{v}=\nu \hat{x}/2k$. 
\label{fig:Two atoms}}
\end{figure}

\emph{Mechanism}.--Two two-level atoms have four quantum
states, a ground state $|gg\rangle$, two first excited states $|ge\rangle$
and $|eg\rangle$, and a double excited state $|ee\rangle$. Considering
the interaction between the two atoms, the
eigen basis of the first excited states is composed by the symmetric and anti-symmetric states,

\begin{equation}
\begin{aligned} & \left|+\right\rangle
=\frac{1}{\sqrt{2}}\left[\left|eg\right\rangle +\left|ge\right\rangle \right],\\
 & \left|-\right\rangle =\frac{1}{\sqrt{2}}\left[\left|eg\right\rangle
-\left|ge\right\rangle \right],
\end{aligned}
\label{eq1}
\end{equation}
with decay rates $\gamma_{\pm}=\gamma_{0}\pm\gamma_{c}$ and energy
shifts $\Delta_{\pm}=\pm\Delta_{c}$. Here $\gamma_{0}$
is the single atom decay rate, $\gamma_{c}$ and $\Delta_{c}$ are the
 collective decay rate and energy shift (see Supplementary Material
\cite{Supplemental}). When the distance between the two atoms $r\ll\lambda$
where $\lambda$ is the transition wavelength, we have
$\gamma_{c}\rightarrow\gamma_{0}$
and thus $\gamma_{+}\rightarrow2\gamma_{0}$ and $\gamma_{-}\rightarrow0$.
The collective energy shift $\Delta_{c}$ is divergent with $1/r^{3}$.
A weak probe field can only resonantly excite $|+\rangle$ from $|gg\rangle$
since the collective energy shift $\Delta_{c}$ moves the transition
between $|+\rangle$ and $|ee\rangle$ out of resonant with the probe
field \cite{Gaetan2009}. We can neglect the two-photon absorption for a weak probe field \cite{Varada1992, Hettich2002}.
The states $|gg\rangle$, $|+\rangle$ and $|-\rangle$ form a three-level
system, as shown in Fig. \ref{fig:Two atoms} (a). The symmetric and the anti-symmetric states satisfy the requirement
on the decoherence rates for EIT, i.e., $\gamma_{+}\gg\gamma_{-}$
when $r\ll\lambda$. The eigenenergies of $|\pm\rangle$ states are
split by the collective energy shift. 

The challenge is how to resonantly couple
$|+\rangle$ and $|-\rangle$ states. The key result of this Letter is that $|+\rangle$ and $|-\rangle$ states can be coupled
by two off-resonant counter-propagating plane waves with different frequencies $\nu_1$ and $\nu_2$. If the frequency
difference $\nu=\nu_1-\nu_2$ matches the splitting between $|+\rangle$ and $|-\rangle$ states $2\Delta_c$, we obtain
on resonance coupling via two Raman transitions as shown in Fig. \ref{fig:Two atoms} (b).
The resulting Hamiltonian is (assuming $\hbar=1$) \cite{Supplemental},
\begin{equation}
\begin{aligned}H= &
\omega_{+}|+\rangle\langle+|+\omega_{-}|-\rangle\langle-|+\Omega_{c}(t)(|+\rangle\langle-|+|-\rangle\langle+|)\\
 & -\Omega_{p}(e^{-i\nu_{p}t}|+\rangle\langle gg|+h.c.),
\end{aligned}
\label{EIT}
\end{equation}
where $\Omega_{c}(t)=\Omega_{0}\sin(kr)\sin(\nu t-\phi)$
with $k=\nu_s/c$, $\nu_s=(\nu_1+\nu_2)/2$, $r=x_1-x_2$ and $\phi=k(x_1+x_2)$ with $x_{1,2}$ being
the coordinates of the two atoms along the propagation of the plane waves. The coupling strength $\Omega_{0}=E^{2}d^{2}/(\omega-\nu_{s})$ with $E$ being the amplitude of the electric field of the plane waves, $d$ being the transition matrix element of the atoms and $\omega$ being the single atom transition frequency. The transition frequencies of $|\pm\rangle$ states are $\omega_{\pm}=\omega\pm\Delta_c+\delta_u(t)$ with $\delta_{u}(t)=\Omega_{0}[1+\cos(kr)\cos(\nu t-\phi)]$ being a universal Stark shift induced by the two plane waves.

The absorption spectra can be calculated by the Liouville equation,
\begin{equation}
\begin{aligned}\frac{\partial\rho}{\partial t}= &-{i}[H,\rho]+\sum\limits
_{j=+,-}\frac{\gamma_{j}}{2}[2|gg\rangle\langle j|\rho|j\rangle\langle gg|\\
 & -|j\rangle\langle j|\rho-\rho|j\rangle\langle j|].
\end{aligned}
\label{me}
\end{equation}
Since $H$ is time-dependent with frequency $\nu$, the coherence
can be expanded $\langle+|\rho|gg\rangle=\sum_{n}\rho_{+gg}^{[n]}e^{in\nu t}$.
Eq.(\ref{me}) can be solved with the Floquet theorem \cite{Agarwal1984,Wang2015} and the
absorption
is proportional to $\text{Im}\rho_{+gg}^{[0]}$, the imaginary part
of the zero frequency coherence.

\begin{figure}
\begin{centering}
\includegraphics[width=7cm]{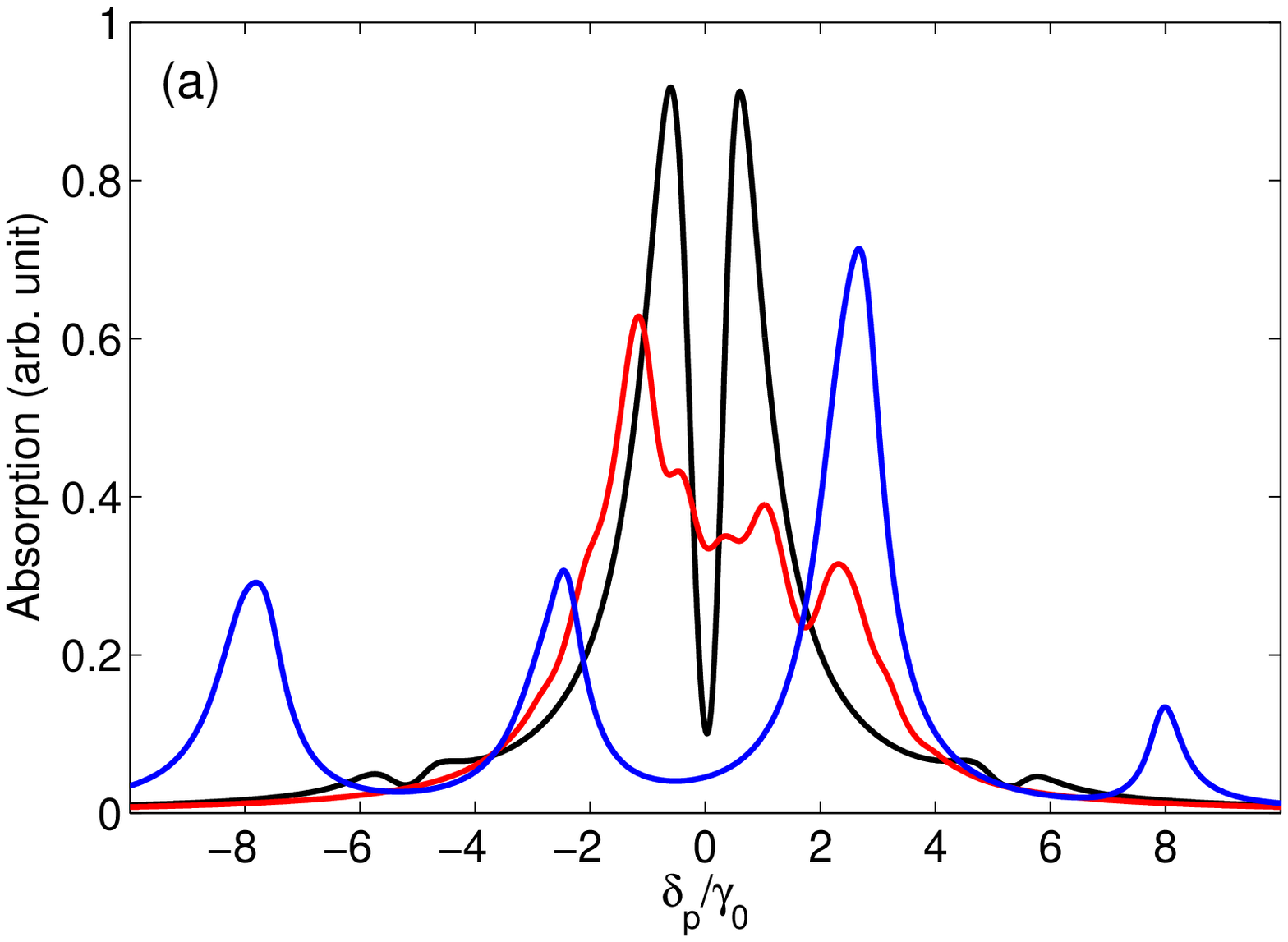}
\par\end{centering}

\begin{centering}
\includegraphics[width=7cm]{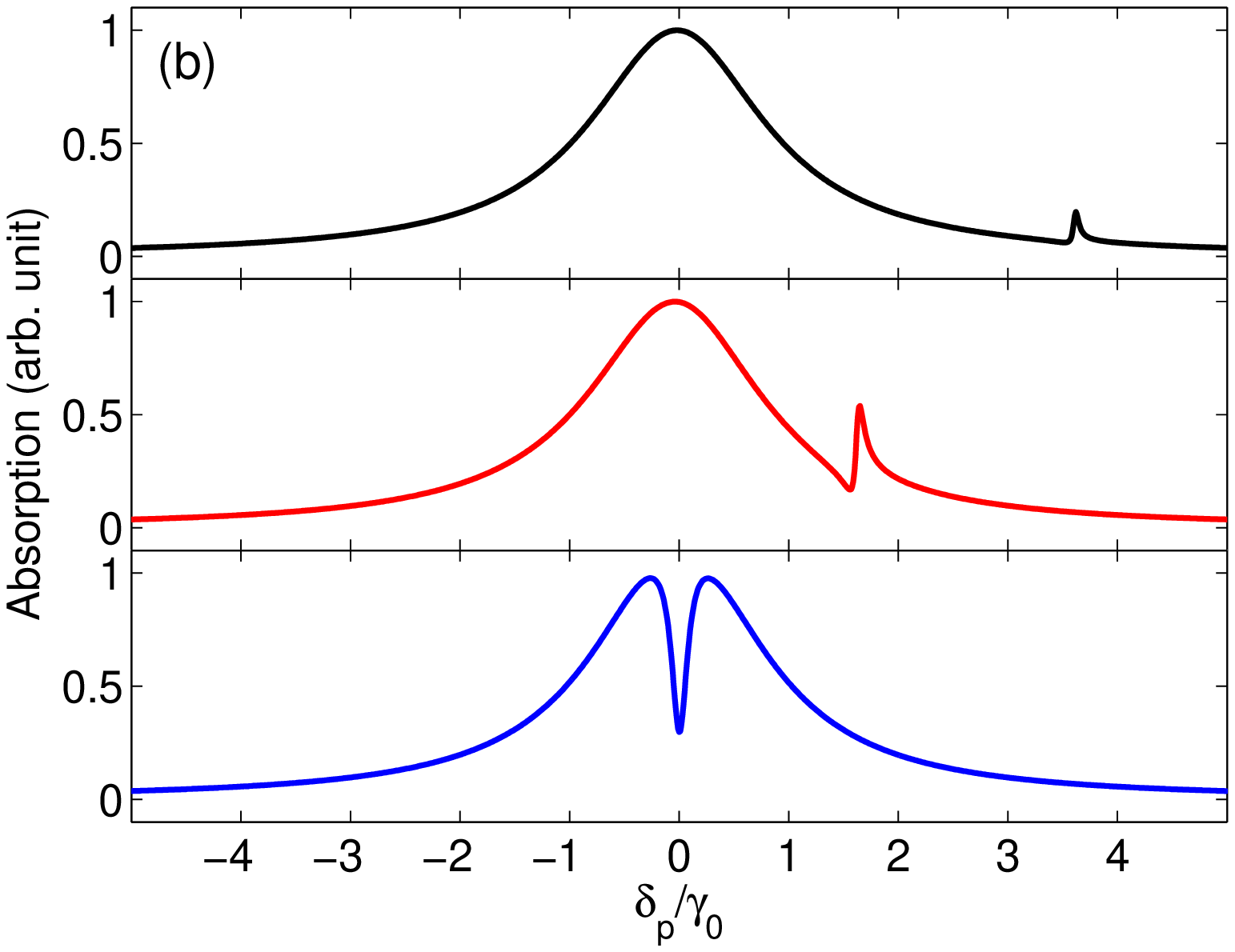}
\par\end{centering}

\centering{}\caption{(Color online) Absorption spectra of two-atom superradiance
EIT. (a) The absorption spectra for different
distances $r$ and Rabi frequencies $\Omega_0$. Black line: $r=0.1\lambda$ ($\Delta_c=2.60\gamma_0$, $\gamma_c=0.92\gamma_0$),
$\Omega_{0}=2\gamma_{0}$; red line: $r=0.2\lambda$ ($\Delta_c=0.38\gamma_0$, $\gamma_c=0.71\gamma_0$), $\Omega_{0}=2\gamma_{0}$
and blue line: $r=0.1\lambda$, $\Omega_{0}=10\gamma_{0}$. The coupling field is
on resonance for each case, $\nu=2\Delta_c$.
(b) The absorption spectra with different standing wave detunings.
$\nu=7\gamma_0$ (black line) $9\gamma_0$ (red line) $10.5\gamma_0$ (blue line).
$\Omega_{0}=\gamma_{0}$. When $\nu=10.5\gamma_{0}=2\Delta_{c}$, the absorption
spectrum is symmetric.
From the relation between $\Delta_{c}$ and $r$, we obtain the
distance between the two atoms $r=0.08\lambda_{0}$, which agrees with
the parameters that we set. 
\label{fig:The-absorption-spectra}}
\end{figure}

The counter-rotating wave terms of $\Omega_c(t)$ can be neglected
for small distance between the two atoms and weak coupling field when
$\Omega_{0}\sin(kr)\ll\Delta_{c}$.
We obtain typical EIT absorption spectra with two absorption peaks
and one transparency point, as shown in the black curve of Fig.
\ref{fig:The-absorption-spectra} (a). Here the probe detuning $\delta_p=\omega+\Delta_c+\Omega_0-\nu_p$ has
taken into account all the static energy shifts of $|+\rangle$ state, including
$\Omega_0$, the static part of the universal Stark shift $\delta_u(t)$. The
effect of the counter-rotating wave terms and the universal shift $\delta_{u}(t)$
emerge either when we increase the distance (reduce $\Delta_c$) between the two atoms
or increase the dynamic Stark shift $\Omega_{0}$ (proportional
to the intensity of the standing wave), which are demonstrated by the multiple side
peaks in Fig. \ref{fig:The-absorption-spectra} (a). 

We can use the following procedure for the subwavelength metrology,
as shown in Fig. \ref{fig:The-absorption-spectra} (b). We first reduce
the intensity of the standing wave to only allow two peaks to appear
in the spectra. Then we tune the frequency difference $\nu$ until
the two absorption peaks become symmetric, which yields the collective
energy shift $\Delta_{c}=\nu/2$. The distance between the two atoms
can be obtained by the relation between $\Delta_{c}(r)$
and $r$ \cite{Supplemental}. Since $\Delta_{c}(r)\propto1/r^{3}$ for small
distance
$r\ll\lambda$, the sensitivity $\delta\Delta_{c}/\delta r\propto1/r^{4}$.
Compared with the existing proposals for subwavelength imaging of two interacting atoms with fluorescences \cite{Chang2006}, a natural preference for this EIT metrology is that both the dressing field and the probe fields are weak. This is in
particular useful for the biological
samples that cannot sustain strong laser fields.

\begin{figure}
\begin{centering}
\includegraphics[width=8cm]{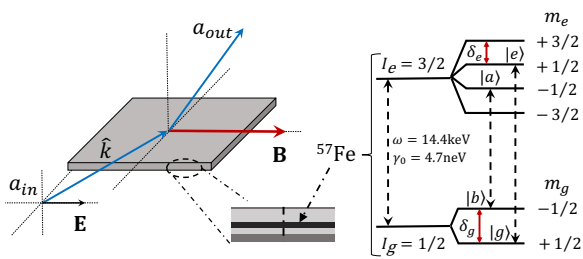}
\par\end{centering}
\caption{Superradiance EIT in nuclear quantum optics. A thin-film cavity
is probed by hard x-rays with grazing angle incidence. The $^{57}$Fe nuclei are embedded
in the center of the cavity. We add an oscillating
magnetic field parallel to the electric field of the linearly polarized incident
x-ray. Only the two transitions
denoted by the dashed arrows between the magnetic Zeeman levels can happen.
The energy difference of these two transitions serve as
the effective coupling between the superradiant and subradiant states.
The EIT spectra can be detected with the reflected signal. \label{fig:Fe}}
\end{figure}

The above mechanism can also be understood as a dynamic modulation of the transition frequency difference between the two atoms \cite{Supplemental}. We notice that the difference
between $|+\rangle$ and $|-\rangle$ states is a relative $\pi$
phase factor between $|eg\rangle$ and $|ge\rangle$ states.
If we can control the transition frequencies of the two atoms
such that the states $|eg\rangle$ and $|ge\rangle$ have energy shifts
$\Omega_{c}$ and $-\Omega_{c}$ respectively, an initial state of the symmetric state
$|\psi(0)\rangle=|+\rangle$ evolves with time
$|\psi(t)\rangle=(e^{-i\Omega_{c}t}|eg\rangle+e^{i\Omega_{c}t}|ge\rangle)/\sqrt{2}$.
At $t=\pi/2\Omega_{c}$, we obtain $|\psi(t)\rangle=-i|-\rangle$. Therefore,
the states $|+\rangle$ and $|-\rangle$ are coupled by an
energy difference between the two atoms. In our scheme, the two counter propagating plane waves create a moving standing wave that induces a time-dependent dynamic Stark shift difference between the two atoms, $\Omega_c(t)$, which serves as the coupling field. This picture enables us to generalize the mechanism to many atoms, as shown later.

The single atom EIT \cite{Mucke2010} and the superradiance and subradiance of
two ions \cite{DeVoe1996} have been observed in experiments. The coupling
between the symmetric and anti-symmetric states has also been realized with two
atoms trapped in an optical lattice \cite{Trotzky2010}. In particular, the
cryogenic fluorescence of two interacting terrylene molecules has been used for
spectroscopy with nanometer resolution \cite{Hettich2002}. Due to different
local electric fields, the two molecules have different transition frequencies,
which corresponds to a static coupling field $\Omega_c$. By introducing an
oscillating electric field gradient or a moving standing wave, such a system can
be exploited for the current EIT experiment of superradiance and subradiance. Very recently, superradiance was also observed from two silicon-vacancy centers embedded in diamond photonic crystal cavities \cite{Sipahigil2016}, which provide another platform to realize this mechanism.

\emph{Generalization to many atoms.}--The mechanism
can be extended to large ensembles of two-level systems. Let us consider
two atomic ensembles, one with $|e\rangle$ and $|g\rangle$, and
the other with $|a\rangle$ and $|b\rangle$ as their excited and
ground states. Each ensemble has $N$ atoms and both ensembles are
spatially mixed together. The transition frequency difference between the two atomic ensembles is within the linewidth such that a single photon 
can excite the two ensembles to a superposition of two timed Dicke
states \cite{Scully2006, Evers2013}, 
\begin{equation}
|+_{\mathbf{k}}\rangle=\frac{1}{\sqrt{2}}(|e_{\mathbf{k}}\rangle+|a_{\mathbf{k}}\rangle)\label{plus}
\end{equation}
where 
\begin{equation}
\begin{aligned} & |e_{\mathbf{k}}\rangle=\frac{1}{\sqrt{N}}\sum\limits
_{n=1}^{N}e^{i\mathbf{k}\cdot\mathbf{r}_{n}}|g_{1},...,e_{n},...,g_{N}\rangle\otimes|b_{1},b_{2},...,b_{N}\rangle,\\
 &
|a_{\mathbf{k}}\rangle=|g_{1},g_{2},...,g_{N}\rangle\otimes\frac{1}{\sqrt{N}}\sum\limits
_{n=1}^{N}e^{i\mathbf{k}\cdot\mathbf{s}_{n}}|b_{1},...,a_{n},...,b_{N}\rangle.
\end{aligned}
\end{equation}
Here $\mathbf{r}_{n}$ and $\mathbf{s}_{n}$ are the positions of
the $n$th atom in the two ensembles. $\mathbf{k}$ is the wave vector
of the single photon. The timed Dicke states $|e_{\mathbf{k}}\rangle$
and $|a_{\mathbf{k}}\rangle$ are excited from the same ground state
$|G\rangle\equiv|g_{1},g_{2},...,g_{N}\rangle\otimes|b_{1},b_{2},...,b_{N}\rangle$
by a single photon. They have directional emission in the direction
of $\mathbf{k}$, so as their superposition state $|+_{\mathbf{k}}\rangle$,
associated with enhanced decay rate and collective Lamb shift. On
the other hand, the state 
\begin{equation}
|-_{\mathbf{k}}\rangle=\frac{1}{\sqrt{2}}(|e_{\mathbf{k}}\rangle-|a_{\mathbf{k}}\rangle),
\end{equation}
is a subradiant state in the sense that its decay rate is estimated to be similar to that of a single atom
\cite{Scully2015PRL}. The directional emissions of $|e_{\mathbf{k}}\rangle$
and $|a_{\mathbf{k}}\rangle$ are canceled because of the relative
phase factor $-1$ between them. The collective Lamb shift of $|-_{\mathbf{k}}\rangle$ can be very
different from that of the $|+_\mathbf{k}\rangle$ state.

We can dynamically couple $|+_\mathbf{k}\rangle$ and $|-_\mathbf{k}\rangle$
states in a well studied nuclear quantum optical system \cite{Rohlsberger2010,
Evers2013, Heeg2013}, as shown in Fig. \ref{fig:Fe}. The nuclei embedded in a
waveguide are $^{57}$Fe with the transition frequency $\omega=14.4\mathrm{keV}$
and the linewidth $\gamma_0=4.7\mathrm{neV}$. In the
presence of a magnetic field, the ground and excited states with
$\textrm{I}_{g}=1/2$
and $\textrm{I}_{e}=3/2$ split into multiplets with Zeeman energy splitting
$\delta_{j}$ ($j=e,g$).
Applying a magnetic field $\mathbf{B}$ parallel to the incident and outgoing
electric fields $\mathbf{E}_\text{in}$ and $\mathbf{E}_\text{out}$ and
perpendicular to $\mathbf{k}$, the linearly polarized
input x-ray can only couple two transitions, as shown in Fig. \ref{fig:Fe}. At
room temperature, the populations on the two magnetic sublevels of the
ground state are approximately equal \cite{Evers2013}. Here we can use a magnetically soft $^{57}$FeNi absorber foil with zero magnetostriction \cite{Tittonen1992} to avoid the mechanical sidebands and other complications in a time-dependent external magnetic field.

The Hamiltonian in the interaction picture can be written as,
\begin{equation}
\begin{aligned}
H 
=&\Omega_{c}\mbox{\ensuremath{\left(t\right)}}(\left|+_{\mathbf{k}}\right\rangle
\left\langle -_{\mathbf{k}}\right|e^{-i\omega_0t}
  +\left|-_{\mathbf{k}}\right\rangle \left\langle
+_{\mathbf{k}}\right|e^{i\omega_0t})\\
  &-\Omega_{p}(e^{-i\delta_{p}t}|+_{\mathbf{k}}\rangle\langle G|+h.c.),
\end{aligned}
\end{equation}
where $\Omega_{c}\mbox{\ensuremath{\left(t\right)}}=\Omega_1\cos\left(\nu
t\right)$ with
$\Omega_1=\left(\text{\ensuremath{\delta}}_{g}+\delta_{e}\right)/2$
is induced by a magnetic field $B=B_{0}\cos\nu t$.
$\omega_{0}$ is the collective Lamb shift difference between the states
$\left|+_{\mathbf{k}}\right\rangle $ and $\left|-_{\mathbf{k}}\right\rangle $.
$\delta_p$ is the probe detuning from the $|+_\mathbf{k}\rangle$ state.
The reflectance of the thin film cavity is dominated by the coherence $|\rho_{+G}|^2$ where $\rho_{+G}\equiv \langle +_{\mathbf{k}}|\rho|G\rangle$ (see \cite{Supplemental}),
\begin{equation}
|R|^2\propto \lim\limits_{T\rightarrow\infty}\frac{1}{T}\int\limits_{0}^{T}|\rho_{+G}(t)|^2
\text{d}t=\sum_{n}\left|\rho_{+G}^{\left[n\right]}\right|^{2},
\end{equation}
where we have made average in a time interval $T\gg 1/\nu$. The
coherence $\rho_{+G}$ has
multiple frequency components $\rho_{+G}(t)=\sum_{n}\rho_{+G}^{[n]}e^{i2\nu t}$ due to the
counter-rotating wave terms. Only when $\nu=0$, no time average is needed.

\begin{figure}
\begin{centering}
\includegraphics[width=7cm]{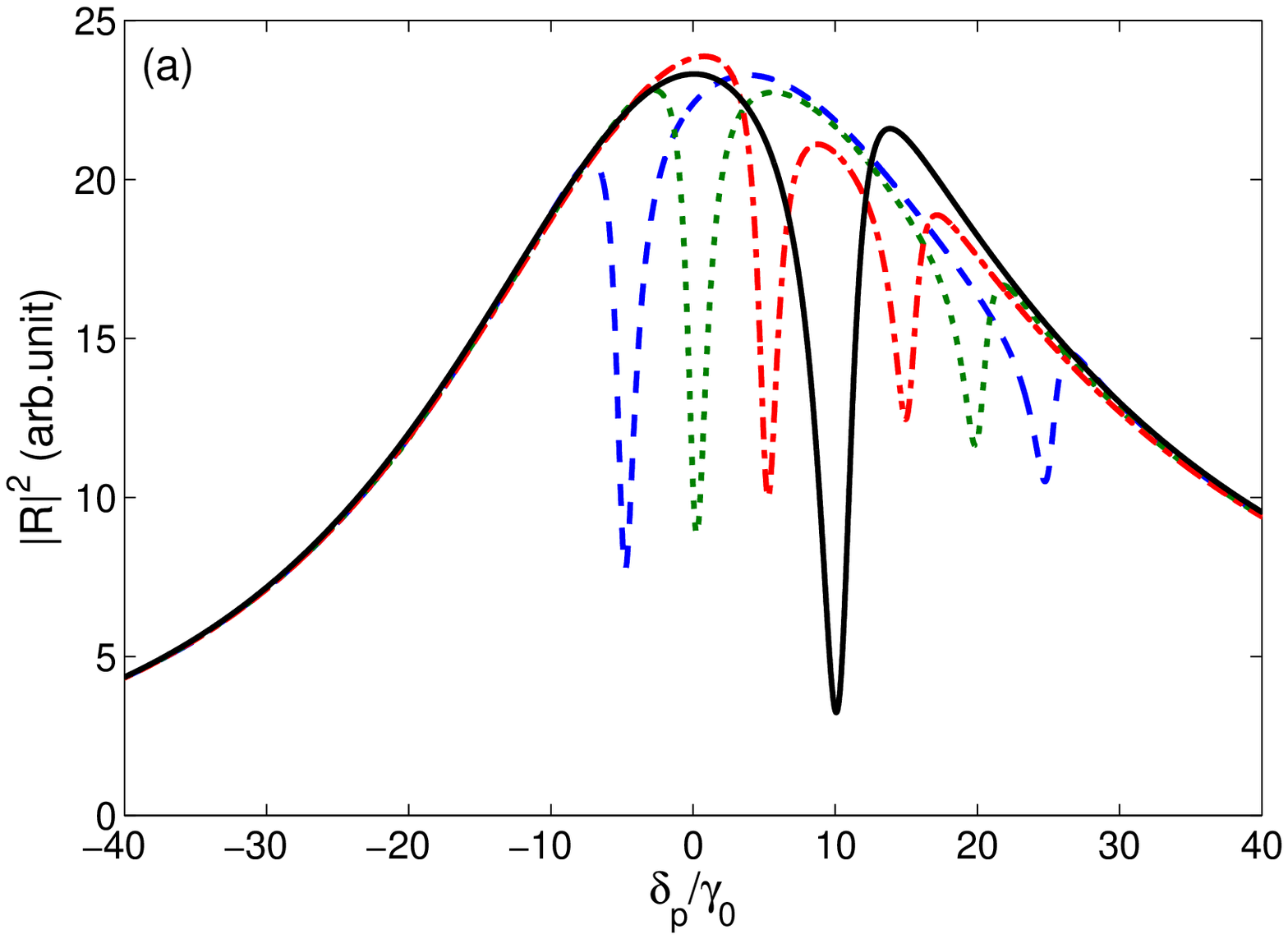}
\par\end{centering}

\begin{centering}
\includegraphics[width=7cm]{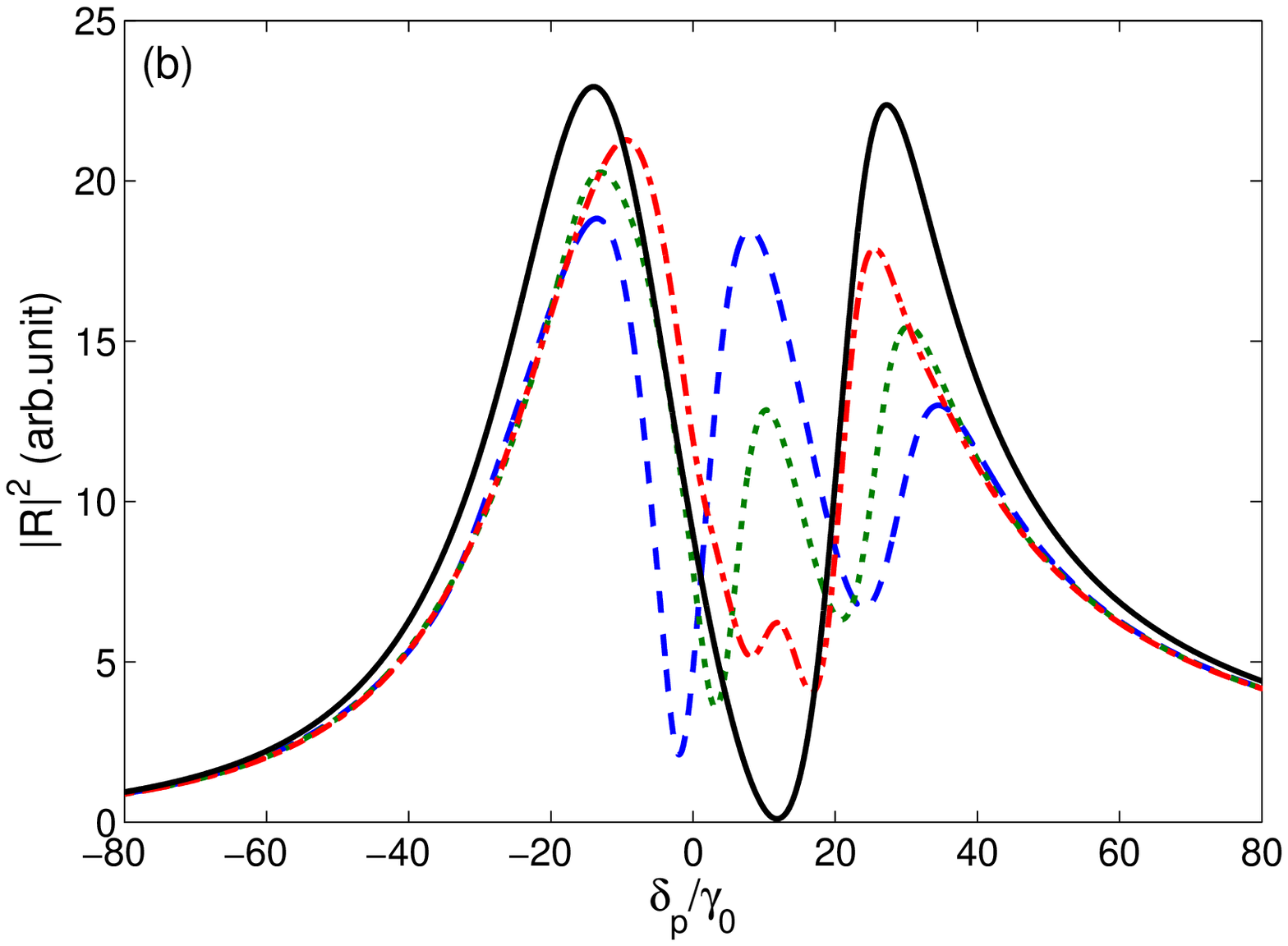}
\par\end{centering}

\caption{(Color online) The reflectance of x-ray with effective Rabi frequencies
(a)
$\Omega_{1}=5\text{\ensuremath{\gamma}}_{0}$ and (b)
$\Omega_{1}=20\text{\ensuremath{\gamma}}_{0}$. The decay rates of
$|+_\mathbf{k}\rangle$ and $|-_\mathbf{k}\rangle$ are $\gamma_+=50\gamma_0$ and
$\gamma_-=\gamma_0$. The collective Lamb shift difference $\omega_0=10\gamma_0$.
The oscillation frequencies of the magnetic fields are $\nu=15\gamma_0$ (blue dash
dot line), $10\gamma_0$ (black solid line), $5\gamma_0$ (red dash line) and $0$
(green dot line). \label{fig:R}}
\end{figure}

The typical collective Lamb shift
of $\mathrm{^{57}Fe}$ nuclear ensemble is $5\sim10\gamma_{0}$
\cite{Rohlsberger2010}.
The internal magnetic field in the $\mathrm{^{57}Fe}$
sample can be tens of Tesla in an external radio-frequency field
\cite{Tittonen1992,FeNuclear}.
The effective coupling field Rabi frequency $\Omega_{1}$
can be easily tuned from zero to $20\gamma_{0}$. The magnetic field
amplitudes corresponding to the effective coupling strengths
$\Omega_{1}=5\gamma_{0}$
and $\Omega_{1}=20\gamma_{0}$ taken in Fig. \ref{fig:R} (a) and (b) are
$B_{0}=5.3\mathrm{T}$
and $B_{0}=21.3\mathrm{T}$, respectively. 

The reflectance spectra can be used to investigate the effect of the
counter-rotating wave terms of the coupling field and to determine the collective
Lamb shift. For a relatively small $\Omega_1$, there are two dips in a single
Lorentzian peak, as shown in Fig. \ref{fig:R} (a). The left and right ones
correspond to the rotating and counter-rotating wave terms of the coupling field,
respectively. The distance between the two dips is approximately $2\nu$. When
$\nu=0$, these two dips merge and the spectrum is the same as the one of the
previous EIT experiments with a static coupling between two ensembles mediated by
a cavity \cite{Rohlsberger2012}. For a larger $\Omega_1=20\gamma_0$ in Fig.
\ref{fig:R} (b), we still have the two dips since $\Omega_1<\gamma_+$ and the
vacuum induced coherence still exists \cite{{Heeg2013}}, but we also have two
peaks basically corresponding to the two magnetic transitions in Fig. \ref{fig:Fe}.
Compared with the result in \cite{Evers2013} where $|+_\mathbf{k}\rangle$ and
$|-_\mathbf{k}\rangle$ have the same energy and the magnetic field is static,
here the two peaks are not symmetric for $\nu=0$ due to a finite Lamb shift
difference. Therefore, the results can be compared with experimental data to
obtain the collective Lamb shifts.

In conclusion, we construct an EIT scheme by dynamically coupling the
superradiant state with the subradiant state. The interaction between atoms can
be measured by the EIT spectra. Compared with the EIT-like schemes with a static
coupling in atomic ensembles \cite{Rohlsberger2012, Evers2013, Heeg2013, Heeg2015,
Makarov2015}, the local dynamical modulation of the transition frequencies of
the atoms introduces a tunable detuning for the coupling field. Therefore, our
scheme contains all the ingredients of EIT. In particular, for the systems where
the splitting between the superradiant and subradiant states is larger than the
decay rate of the superradiant state, the dynamic modulation can bring the EIT
dip to the Lorentzian absorption peak of the superradiant state, as shown in
Fig. \ref{fig:The-absorption-spectra} (b). The dynamic modulation enables a
precise measurement of the distance between two atoms and brings new physics of
the EIT point due to counter-rotating wave terms.

The authors thank G. Agarwal, A. Akimov, A. Belyanin, J. Evers, O.
Kocharovskaya, R. R\"ohlsberger and A. Sokolov for insightful discussions. We acknowledge the support of National Science Foundation Grant EEC-0540832 (MIRTHE ERC), Office of Naval Research (Award No. N00014-16-1-3054) and Robert A. Welch Foundation (Grant No. A-1261). Wei Feng was supported by China Scholarship Council (Grant No. 201504890006). H. Cai is supported by the Herman F. Heep and Minnie Belle Heep Texas A$\&$M University Endowed Fund held/administered by the Texas A$\&$M Foundation.

\end{document}